# Single-shot measurement of frequency-resolved state of polarization dynamics in ultrafast lasers using dispersed division-of-amplitude

## Author Information

### Affiliations

**[1] Key Laboratory of Optoelectronic Technology & Systems (Ministry of Education), Chongqing University, Chongqing 400044, China**

**[2] Dipartimento di Ingegneria dell'Informazione, Elettronica e Telecomunicazioni, Sapienza Università di Roma, via Eudossiana 18, 00184 Roma, Italy.**

**[3] Novosibirsk State University, 1 Pirogova str, Novosibirsk 630090, Russia.**

QIANG WU[1] , LEI GAO[1,*] , YULONG CAO[1] , STEFAN WABNITZ[2,3] , ZHENGHU CHANG[1] , AI LIU[1] , JINGSHENG HUANG[1] , TAO ZHU[1,*]

### Contributions

Q. W. conducted the experiments. Y. C. and A. L. and Z. C. built the fiber laser systems. Q. W. and J. H. captured the data. Q. W. processed the data and wrote the manuscript. L. G. and S. W. revised the manuscript. L. G. and T. Z. supervised the project. All authors discussed the results.

### Corresponding author

Correspondence to: gaolei@cqu.edu.cn, zhutao@cqu.edu.cn

## Abstract

Precise measurement of multi-parameters of ultrafast lasers is vital both in scientific investigations and technical applications, such as, optical field manipulation, pulse shaping, sample characteristics test, and biomedical imaging. Tremendous progress in parameter measurement of ultrafast laser has been made, including single-shot spectra acquired by time-stretch dispersive Fourier transform in spectral domain, and pulse magnification or compression realized by time lens in temporal domain. Nevertheless, single-shot measurement of frequency-resolved states of polarization (SOPs) of ultrafast lasers has not been reported so far, and the unregular SOP evolution dynamics in ultrafast pulses is hardly explored. Here, we demonstrate a new single-shot frequency-resolved SOPs measurement system by utilizing division-of-amplitude method under far-field approximation. Large dispersion is utilized to time-stretch the laser pulses, where the spectrum information is mapped into temporal waveform via dispersive Fourier transform. By calibrating system matrix with different wavelengths, the precise frequency-resolved SOPs are obtained together with high speed opto-electron detection. We demonstrate applications in direct measurement of transient mode-locked fiber laser dynamics. We observe complex frequency-dependent SOPs dynamics in the building up of dissipative solitons, and apparent discrepancy of SOPs between sideband and main peak in conventional solitons. Our observations reveal that the SOP plays a far more complex part in mode-locking process, which is different from the traditional viewpoint. Taking advantage of broadband achromatic optical elements, this method can be extended to measurement of much broad pulse lasers, which will pave the way for reliable measurement and precise control of ultrafast lasers with frequency-resolved SOPs structures.

## Introduction

Characterization of state of polarization (SOP) of ultrafast laser play an essential part in various fields ranging from fundamental science to applications, such as, investigation of instantaneous evolution dynamics of laser, optical field manipulation and synthesis, pulse shaping, and polarization imaging. These years, much efforts have devoted to temporal-spectral characteristics[1,2]. Based on time-stretched dispersive Fourier transform (DFT) technology, single-shot real-time measurement of ultrafast laser have been achieved, including transient buildup dynamics and interaction of ultrafast pulses[3-12]. In addition to the spectrum measurement under temporal-spectral mapping of DFT, much work with respect to temporal measurement of ultrafast lasers are reported. Time-lens, which is the counterpart of space lens in temporal domain, allows for the single-shot characterization of temporal waveforms via expansion or compression[13-21]. However, to the best of our knowledge, there has been rare work investigating the single-shot measurement of frequency-resolved SOP of ultrafast laser.

Actually, Stokes parameters are frequently utilized to represent SOP owing to simple form with real number but complete characterization of arbitrary SOP nonetheless. Therefore, frequency-resolved SOP of ultrafast laser needs to be done in the single-shot measurement of frequency-resolved Stokes parameters. Although vectorial characteristics are considered in the research of ultrafast fiber lasers before[22-33], key point of which is still the single-shot spectra through orthogonal decomposition or projection instead of Stokes parameters. And two existing work deal with frequency-resolved SOP by filtering[34,35], but not with single-shot measurement. Up to now, two methods have been used to measure the stokes parameters. One is time-sharing method like rotating wave plate[36,37], which is just suitable for constant or slow-varying signals as continuous wave. The other method is beam splitting. Take division-of-amplitude for example, fast changing signals of four channels can be detected simultaneously by this method[38-40]. And spectral polarization measurements can be achieved by employing the grating[41]. Unfortunately, SOP of ultrafast laser still cannot be measured only by division-of-amplitude, on account of the insufficient response speed of conventional instrument. Hence, time-stretching and frequency-resolved techniques are the crucial part.

In our work, we measure SOP of ultrafast laser pulses based on the division-of-amplitude. Taking advantage of this method, we can detect rapidly varying signals of four channels simultaneously. Furthermore, in order to capture the ultrafast pulses with frequency-resolved SOP, DFT technique is taken into account. Combining the two means, the single-shot optical spectra can be measured by high-speed photodetector and displayed on digital oscilloscope, and frequency-resolved SOP is available through the frequency-resolved system matrix. In this case, two types of ultrafast laser pulses are selected as signals under test. Dissipative soliton (DS) with broad and flat-top spectrum is easily recognized and frequency-dependent SOP is obvious feature. Conventional soliton (CS) with formation of Kelly sideband after beating dynamics is also fascinating to us. This method paves a new way to measure the SOP of ultrafast lasers. Besides, SOP offers abundant information for us to understand intricate dynamic mechanism, manipulation of ultrafast laser and full-Stokes imaging[42].

## Results

### Principle and experimental setup

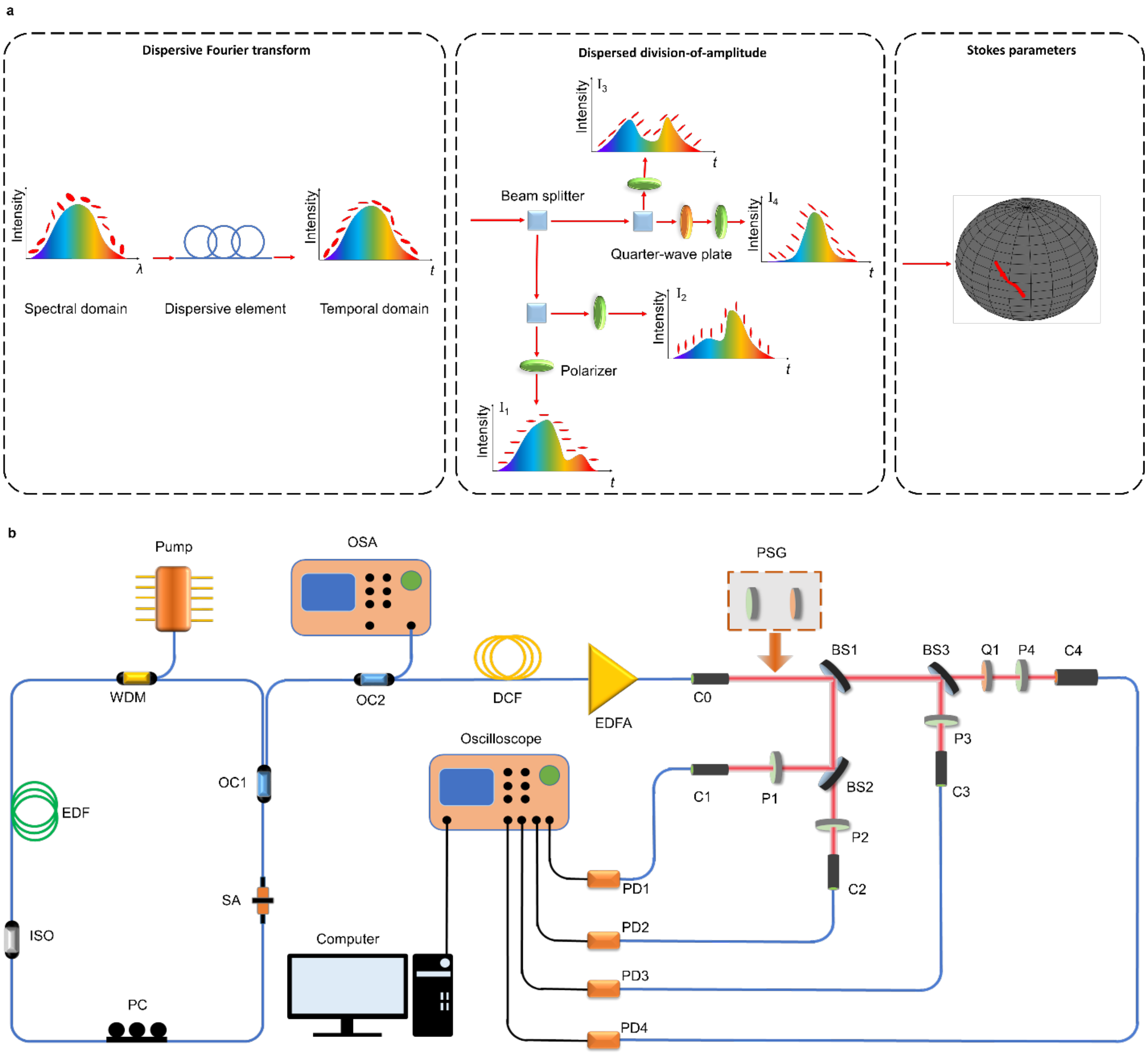

**Fig. 1.a** Principle of DFT and dispersed division-of-amplitude. Ultrafast pulse is time-stretched and the optical spectrum is mapped into temporal domain, under the far-field approximation exerted by dispersive medium. For the dispersed SOP (red ellipses) beam-splitted by beam-splitting element, the output SOP is converted and the intensity distribution of four channels vary totally different. **b** Schematic of the fiber laser cavity as a laser source, including DS and CS, and measurement system containing dispersive element of dispersive compensation fiber and spatial division-of-amplitude. The blue lines denote the fiber system and red lines indicate spatial laser beam, and black lines are electrical lines.

For the measurement of Stokes parameters, owing to the birefringence of polarization-conversion devices, the system is frequency-dependent. So optical spectrum of signal to be measured is necessary. Unfortunately, by virtue of lower response speed, existing optical spectrum analyzers offer averaged optical spectrum, instead of single-shot measurement. In this case, DFT technique is adopted as in Fig.1**a**. Dispersive element is made up with dispersive compensation fiber. Due to the large group velocity dispersion, the propagation of input pulse in the dispersive element is under the far-field approximation in temporal domain, so the spectrum is mapped into a temporal waveform. In this case, higher-order dispersion coefficients are negligible in the one-to-one mapping. And the relation between optical spectrum and temporal waveform is as follows[3]:

$$\Delta t = |D| L \Delta\lambda \tag{1}$$

where $\Delta t$ is the time duration after frequency-to-time mapping, $D$ is the group velocity dispersion of dispersive medium, $L$ is the propagation distance, $\Delta\lambda$ is the optical spectrum bandwidth of the pulse laser. Unlike the averaged processing of commercial optical spectrum analyzer, DFT facilitates single-shot measurement of optical spectrum of ultrafast pulse trains.

Time-stretched pulses with dispersed SOP are beam-splitted when injected into system of division-of-amplitude[35,36]. For each wavelength, the SOP is described by Stokes vector $S = (S_0, S_1, S_2, S_3)^T$ and output laser is represented by intensity vector with four parameters $I = (I_0, I_1, I_2, I_3)^T$ as depicted in Fig.1**a**. Because intensity detected by photodetector in each channel can be expressed as the linear combination of four Stokes vector $I_i = a_{i1}S_0 + a_{i2}S_1 + a_{i3}S_2 + a_{i4}S_3$ .We use a $4\times 4$ system matrix to express the relationships between the intensities and the Stokes parameters $I = AS$. when determinant of system matrix is not equal to zero, viz., $\det(A) \neq 0$, then the inverse matrix exists:

$$S = A^{-1}I \tag{2}$$

And system matrix $A$ is determined by the Stokes vector of each wavelength of input laser and the output intensity. When the intensities of four channels of ultrafast pulse laser are measured, the SOPs can be calculated by the intensities and the inverse matrix $A^{-1}$ of each wavelength. Furthermore, the spherical orientation angle $\theta$ and ellipticity angle $\psi$ of ellipse can be calculated from the Stokes parameters, $\theta = \frac{1}{2}\arctan\left(\frac{S_2}{S_1}\right)$, $\psi = \frac{1}{2}\arctan\left(\frac{S_3}{\sqrt{S_1^2 + S_2^2}}\right)$. And the deviation of SOPs is defined as the relative distance between two points on Poincaré sphere:

$$\Delta S = \sqrt{\left(S_1' - S_1\right)^2 + \left(S_2' - S_2\right)^2 + \left(S_3' - S_3\right)^2} \tag{3}$$

As schematically depicted in Fig.1**b**, we use two laser source with similar configuration. The DS laser ring cavity contains 15 m erbium-doped fiber (EDF) with the dispersion of -12.2 ps/nm/km, forward pumped by a 976 nm diode laser through a 1550/980 nm wavelength division multiplexer (WDM). The saturable absorber (SA) is made from single-wall carbon nanotubes. The rest of the cavity includes an isolator (ISO), a polarization controller (PC), 2.7 m single mode fiber with the dispersion of 18 ps/nm/km, and an optical coupler (OC1) with a 10% output port. The net normal dispersion is 0.171 ps$^2$. The dispersive Fourier transform is achieved by using a 2 km dispersion compensating fiber (DCF) with the dispersion about 300 ps/nm/km. We adopt erbium-doped fiber amplifier (EDFA, Amonics AEDFA-23-B-FA) to get a higher power because of loss of DCF and spatial system. And optical spectrum of the output signal is measured by optical spectrum analyzer (OSA, Yokogawa AQ6370). At the same time, for the CS fiber laser cavity, the SA is also constituted by home-made single-wall carbon nanotubes. By tuning the length of fibers, the net dispersion of conventional soliton is about -0.27 $ps^2$.

For the part of division-of-amplitude, we use spatial system containing a collimator (C0) to get a collimating beam, which is divided into four paths by three beam splitters (BS1-BS3) with transmission reflection ratio of 5:5. Angles of four analyzers (P1-P4) are set as 0°, 45°, 90°, 135°, and the intersection angle between quarter-wave plate (Q1) and analyzer in the fourth path is 45°. The orientation of 0° is parallel to platform. In this way, any state of polarization including linear polarization and circular polarization can be recognized by this system. In order to input arbitrary state of polarization in calibration process, a polarizer and a quarter-wave plate play the role of polarization state generator, which is shown on the pink plate and marked by a pink arrow. The four channels of output laser are received by four collimators (C1-C4). Photoelectric conversion is achieved by four high-speed photodetectors (PD1-PD4) with analog bandwidth of 8 GHz and a digital oscilloscope (Tektronix, DSA 72004B) with the bandwidth of 20 GHz.

## Calibration of measuring system

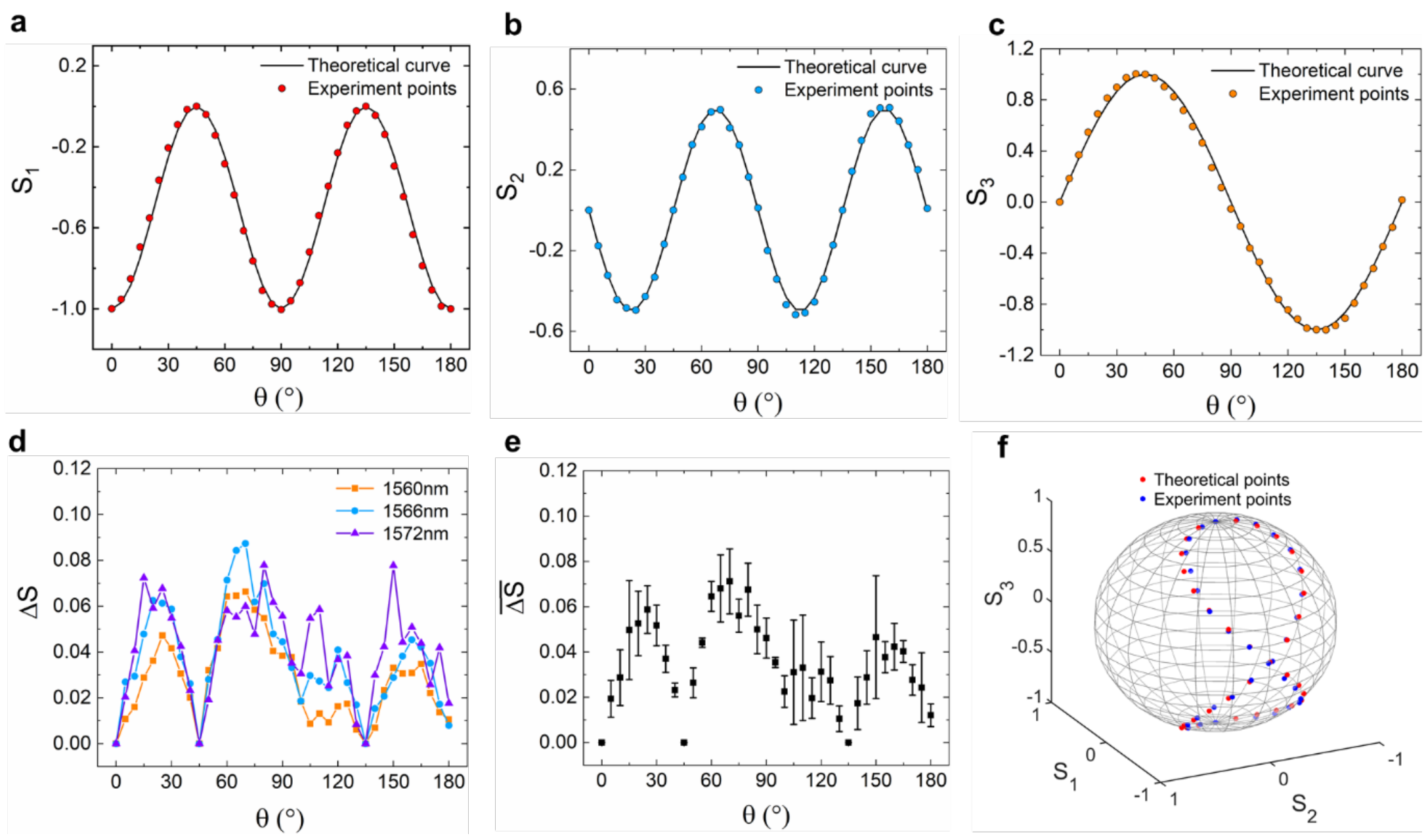

**Fig. 2.** Calibration of division-of-amplitude. **a-c** theoretical value (black solid curve) and experiment points (colored circles) of three stokes parameters at 1550 nm. The polarizer is set at the angle of 90°, and the rotation angle of quarter-wave plate is 180°. So the rotation period of three Stokes parameters are 90°, 90° and 180°, respectively. **d-e** Deviation and mean error of Stokes parameters at different wavelengths of 1560 nm, 1566 nm and 1572 nm, which almost covers the spectrum bandwidth that we used. And the mean deviation is less than 0.07 for three wavelengths. **f** theoretical points (red circles) and measured SOP (blue circles) located on Poincaré sphere. The trace forms a figure-eight pattern from north pole to south pole on Poincaré sphere, which depends on the combination of angle of polarizer and quarter-wave plate.

Before measurement, the system of division-of-amplitude was calibrated at first. In the calibration process, we use tunable semiconductor laser as source and connected with the first collimator. Then we fixed the polarizer at the angle of 90° and rotated the quarter-wave plate from the intersection angle of 0° to 180°, by the step of 5°. And the 180° is variation period. There are two methods of calibration. One is method of four-points. Four points on Poincaré sphere construct a tetrahedron with a nonzero volume, which is equivalent to nonzero determinant or that invertible matrix exists. The other method is equator-poles. We choose four points located at the equator and poles as four known vectors to do the calibration because this method can reduce the effect induced by imperfection of optical elements.

Fig.2 depicts the calibration results of equator-poles. Four known vectors of state of polarization are $(1,-1,0,0)^T$, $(1,0,1,0)^T$, $(1,0,0,1)^T$, $(1,0,0,-1)^T$. As shown in Fig.2**a-c**, measured $S_1$, $S_2$ and $S_3$ at 1550 nm are almost distributed on theoretical curves. In Fig.2**f**, measured points on Poincaré sphere are located near theoretical points with mean deviation of about 0.03. And the distributed trajectory of points is figure-8, but the pattern and location depend on the rotation angle of polarizer and quarter-wave plate. After calibration at 1550 nm, we test the division-of-amplitude system at a wide range of wavelength. As displayed in Fig.2**d-e**, selected wavelengths are 1560 nm, 1566 nm and 1572 nm. Deviation of state of polarization are roughly the same at three wavelengths and mean value of deviation for every rotation angle at three wavelengths is under 0.07. From the result of calibration, the constructed division-of-amplitude system can measure the SOPs at least for the range of a dozen nanometers. For a broader range of wavelengths, normal quarter-wave plate can be replaced by broadband achromatic quarter-wave plate with several hundred or even more nanometers.

## SOPs of dissipative soliton fiber laser

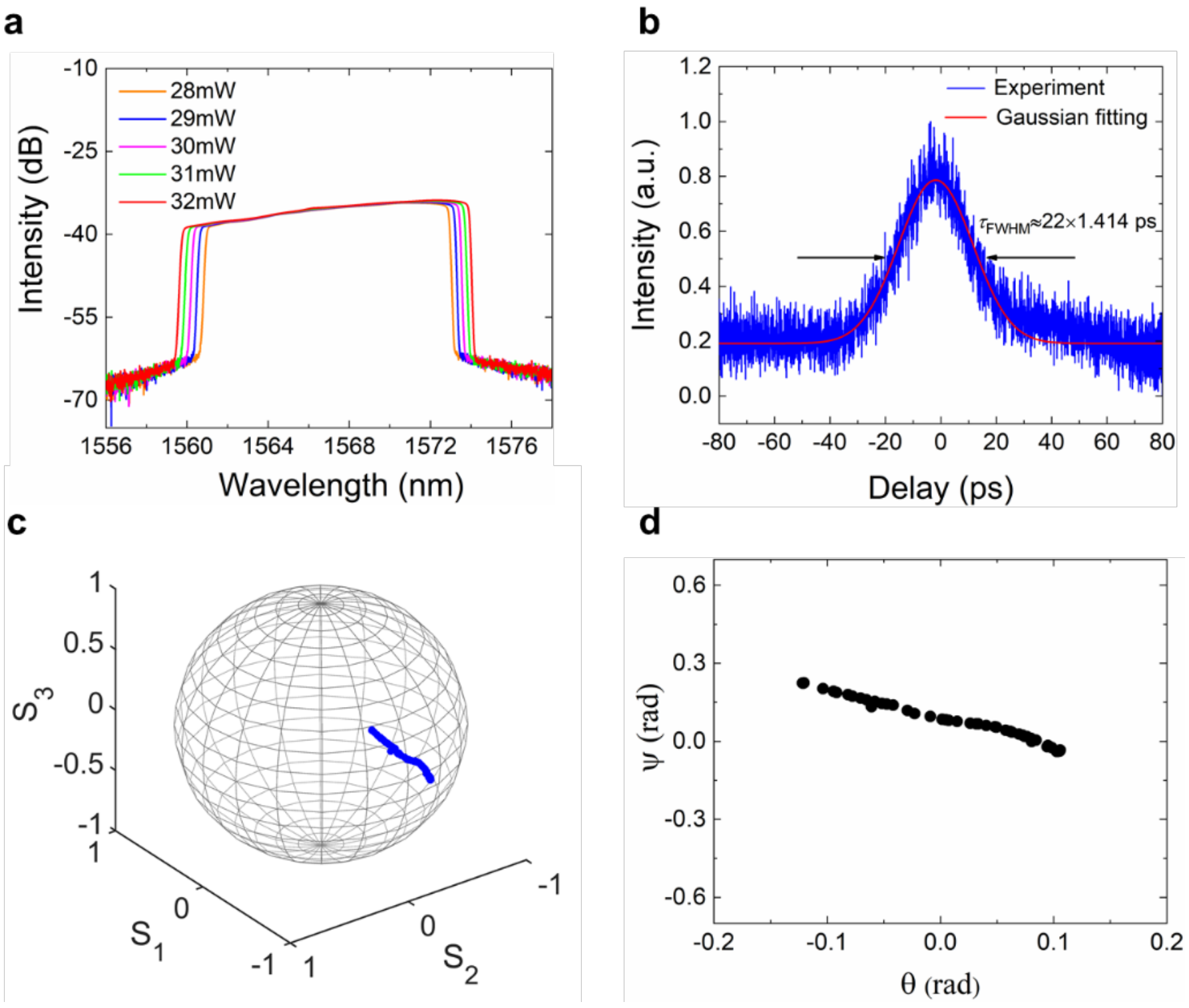

**Fig. 3.** Characteristics of dissipative soliton fiber laser cavity. **a** Optical spectrum under different pump powers ranging from 28 mW to 32 mW. And the bandwidth is about 12 nm to 14 nm. **b** Autocorrelation trace of DS and the fitting curve achieved by Gaussian function. The pulse duration is about 22 ps. **c** SOPs qualitively measured for various filtered wavelengths. And the trace is smooth along one direction. **d** Phase diagram based on the spherical orientation angle $\theta$ and ellipticity angle $\psi$ calculated from the SOPs in **c.** The spherical orientation angle $\theta$ changes about 0.25 radian and the ellipticity angle $\psi$ varies nearly 0.3 radian, respectively.

As shown in Fig.3**a**, the averaged optical spectrum is ranging from 1561 nm to 1573 nm and the intensity is about -35 dB at the pump power of 28 mW, which is measured by optical spectrum analyzer. With the increasing pump power, the spectrum is broadened accordingly. In this process, lower peak power and reduced nonlinear effect enable DS to be self-consistent, due to wave-breaking-free. In Fig.3**b**, we measure the autocorrelation trace by using an autocorrelator (APE, Pulse check USB 150) and EDFA for amplified average power of 3 mW. The autocorrelation trace of DS is fitted by Gaussian function and full width at half maximum of pulse duration is about 30 ps. Furthermore, as depicted in Fig.3**c**, we filtered the DS by using a tunable optical filter (Santec, OTF-320). SOPs for various filtered wavelengths are measured by a polarization state analyzer (General Photonics, PSGA-101-A). As we can see, there is not a fixed point on Poincaré sphere, so SOPs for different wavelength component are different from each other. From the blue side to the red side of the spectrum, a smooth trajectory appears on Poincaré sphere. According to the SOPs calculated above, phase diagrams based on the spherical orientation angle $\theta$ and ellipticity angle $\psi$ are illustrated in Fig.3**d**. Black dots correspond to different wavelengths. The trace of the points in phase diagram is similar to that on Poincaré sphere. The spherical orientation angle $\theta$ changes about 0.25 radian and the ellipticity angle $\psi$ varies nearly 0.3 radian.

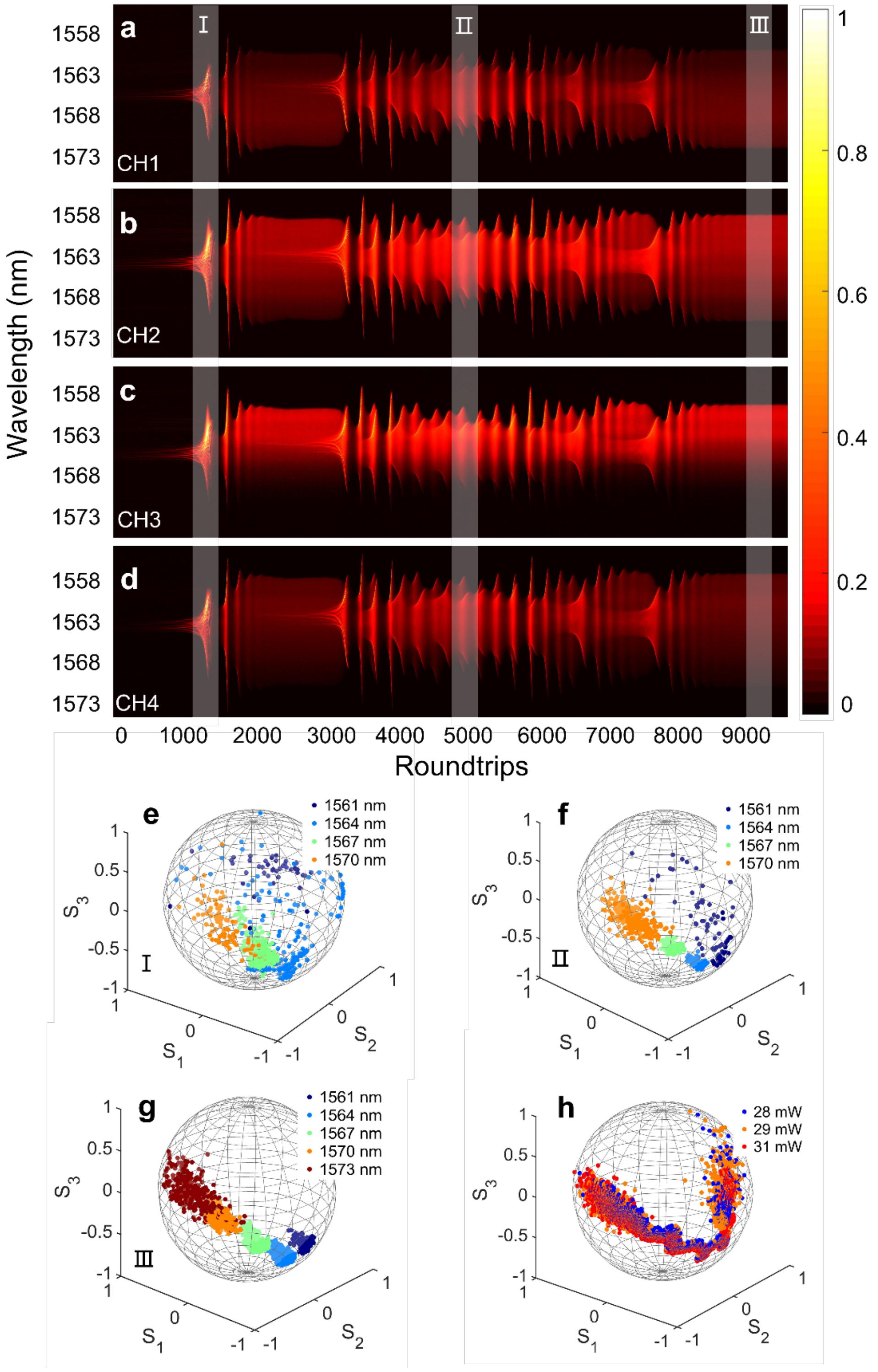

**Fig. 4. a-d** single-shot spectra of four channels (CH1-CH4) under the pump power of 28 mW. Three regions I, II and III in buildup process ranging from relaxation oscillation to stable mode-locking pulses are marked by white bars, which correspond to 1100$^{th}$-1400$^{th}$, 4800$^{th}$-

5100$^{th}$, 9000$^{th}$-9300$^{th}$, respectively. **e-g** SOPs for three regions of DSs, corresponding to white bars in **a-d**, colored by different wavelengths. Selected wavelengths are 1561 nm, 1564 nm, 1567 nm, 1570 nm and 1573 nm. From region I to region III, SOP evolve from chaotic pattern all the way to concentrated and ordered island. **h** SOP in region III under the pump power of 28 mW, 29 mW and 31 mW. The trajectories of three pump powers are similar as a whole, and SOP in the middle overlap. But for the wavelength at the edge, SOP changes sharply and tiny variation can not be recognized in this case.

The single-shot spectra of DS build-up process in every channels detected by high speed photodetector are shown in Fig.4**a-d**, under the pump power of 28 mW. There are more than 9000 roundtrips with time period of 134ns and single DS pulse is time-stretched to 7.2 ns through DFT. We improve the signal-to-noise ratio of every channel by using EDFA before division-of-amplitude to obtain amplified energy. In our measurement system, firstly, detection devices of all channels can be almost the same, and secondly due to the DFT, DS pulses are time-stretched to magnitude of nanosecond, so the tiny difference of arrival time among pulses of four channels can be easily compensated by tuning the optical path in spatial part or by algorithm in data post-processing. When measuring system is applied in real-time data acquisition and processing, the optical path must be precisely adjusted so that the time difference can be ignored, due to limited temporal resolution of detection devices. According to the intensities for various wavelengths of four channels and the inverse matrix of the division-of-amplitude system, we calculate standard normalized stokes parameters of them. The region I, II and III are chosen in two typical stages in buildup dynamics. For region I, relaxation oscillation begins with process of population inversion. Central frequencies grow with energy accumulation. As shown in Fig.4**e**, the corresponding distribution of SOPs is a bit of chaotic, especially for the wavelengths on the edge. In region II, broadened spectrum with periodic sharp peak and bandwidth oscillation, more frequencies appear and SOPs tend to be more concentrated on Poincaré sphere in Fig.4**f**. When evolved into region III, mode-locking mechanism dominates and stable DSs are constructed. As depicted in Fig.4**g**, for the stable DSs, distribution of SOPs on Poincaré sphere is almost a slender island and slowly varies from short wavelength side to long wavelength side, which is in a more ordered state. For increasing pump powers in region III, three of them are selected to display in Fig.4**h**. The main traces are the same under these pump powers, but SOP at the edge varies so intensely that the details cannot be resolved precisely. Deviation among pulses attributes to the jitter or fluctuation of intensities induced by the electronic devices, particularly in measurement of high-frequency signal.

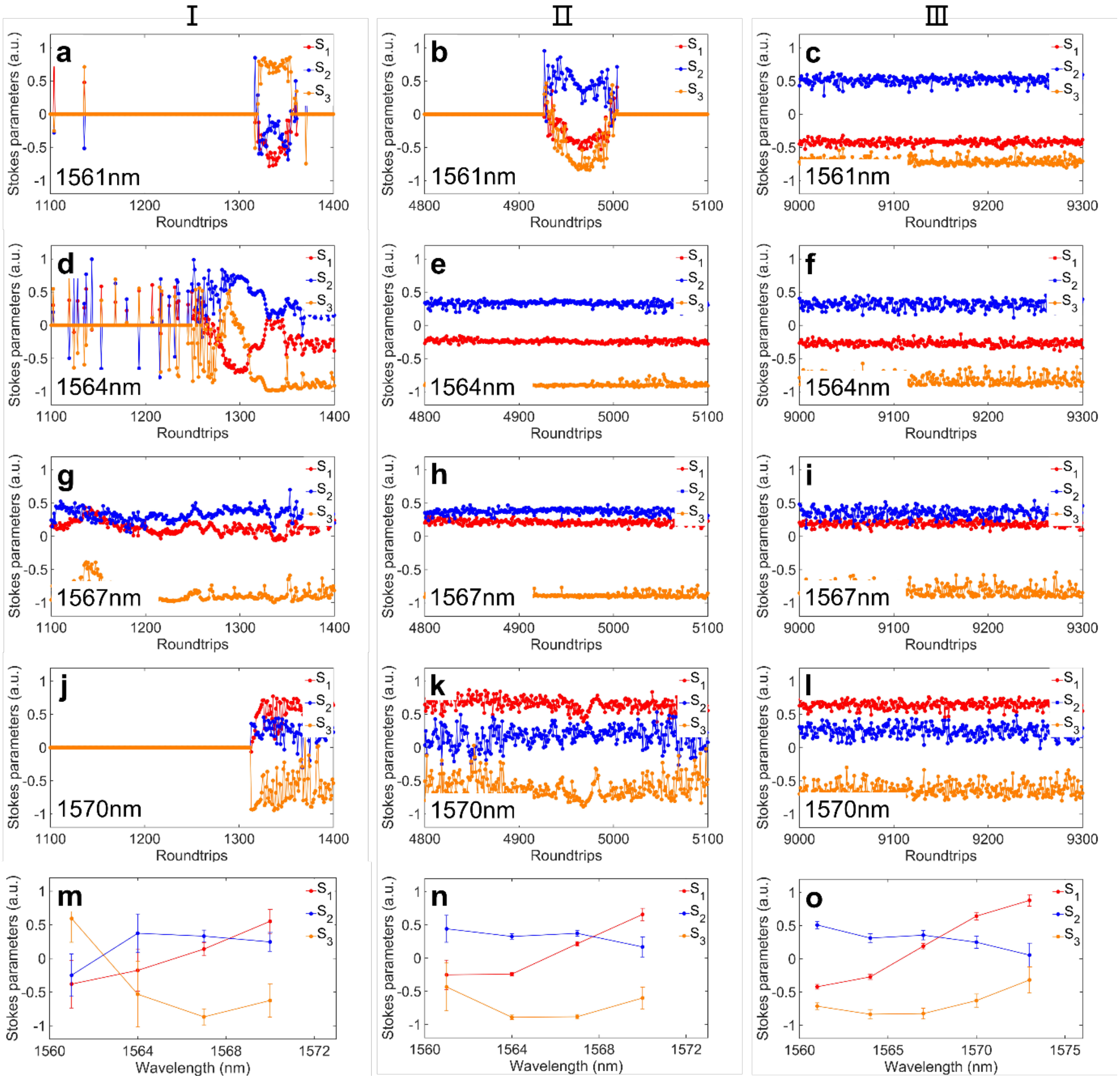

**Fig. 5. a-l** Stokes parameters evolve with roundtrips for different wavelengths of 1561 nm, 1564 nm, 1567 nm and 1570 nm in three regions I, II and III in Fig. 4. For wavelength of 1561 nm in region I and II, Stokes parameters vary intensely in relaxation oscillation. And in mode-locking state, Stokes parameters reach a constant. For 1564 nm and 1570 nm in region I, as in Fig.5**d** and **j**, intense oscillation and polarization conversion occur during the process of bandwidth broadening and Stokes parameters are constant in region II and III. For 1567 nm, Stokes parameters change slightly. **m-o** Mean value and standard deviation of Stokes

parameters with wavelengths in three regions. Large deviation is demonstrated in region I, which is apparent compared with Fig.5**n** and **o.**

Besides the evolution of SOPs with wavelengths on Poincaré sphere, Stokes parameters varying with roundtrips are displayed in Fig.5. For wavelength of 1561 nm in region I and II, Stokes parameters experience local large fluctuations because of sharp peak appearance on the edge of spectrum in relaxation oscillation (Fig.5**a** and **b**). And in region III of mode-locking regime, Stokes parameters maintain a steady level (Fig.5**c**). It should be pointed out that all Stokes parameters with zero values does not mean unpolarized light is measured. In data processing, we just remove points when all intensities of four channels as low as noise in order to eliminate effect of electronic noise, because the varying intensities is not mainly induced by state of polarization of input laser in this case. Therefore, these points cannot be used to calculate SOPs and Stokes parameters are replaced by zero values. For 1564 nm and 1570 nm in region I (Fig.5**d** and **j**), intense oscillation and polarization conversion appear accompanied by bandwidth broadening. Then Stokes parameters remain stable in region II and III. For 1567 nm near the central wavelength, Stokes parameters change slightly all the way, which means SOPs are slightly affected in the process. In addition, as depicted in Fig.5**m-o**, mean values and standard deviation of Stokes parameters with different wavelengths are calculated over the roundtrips in region I, II and III. The whole tendency and the range of error bar are the same as mentioned above.

## SOPs of conventional soliton fiber laser

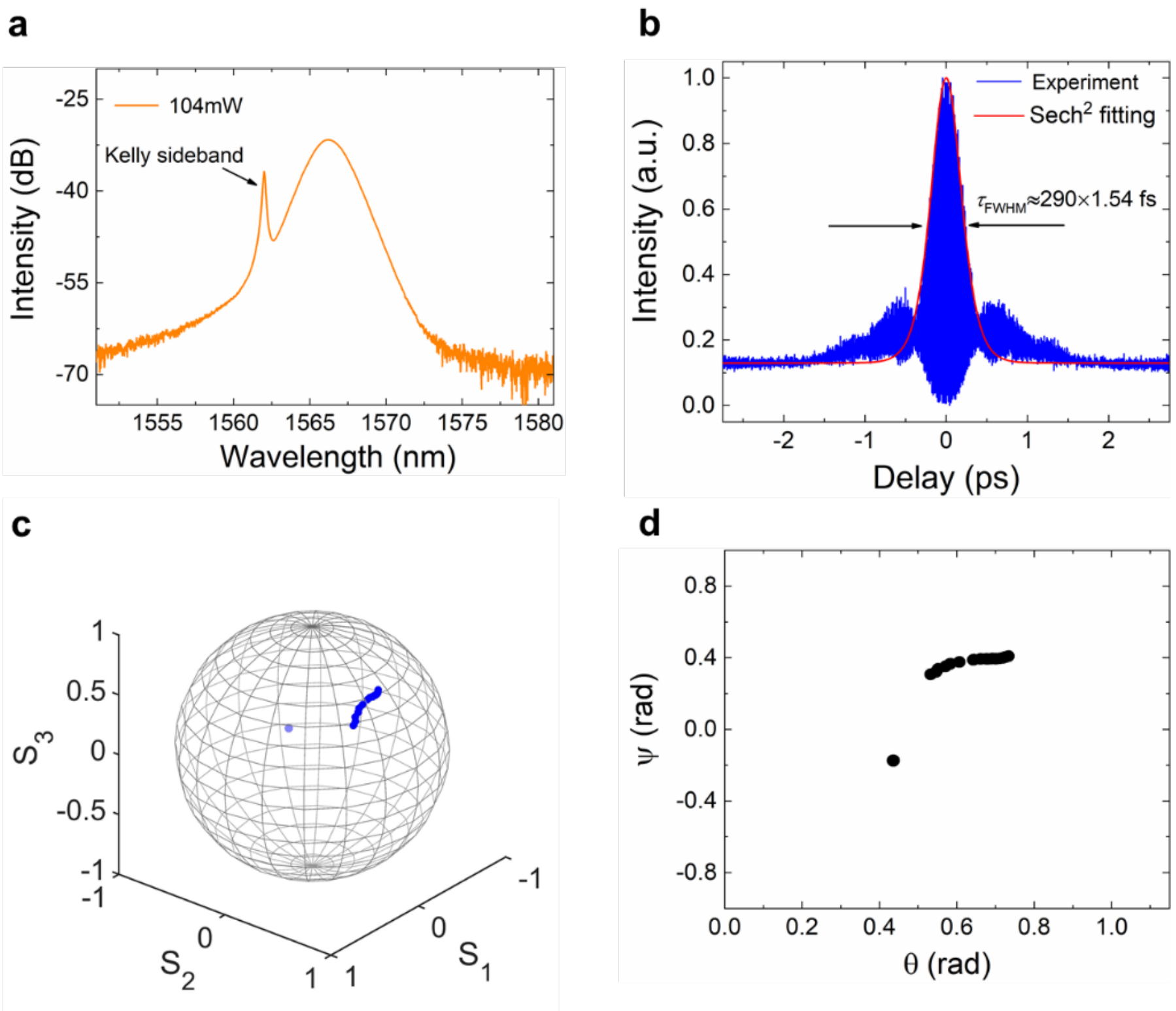

**Fig. 6.** Characteristics of CS fiber lasers. **a** Averaged optical spectrum with one Kelly sideband under the pump power of 104 mW. The 3dB bandwidth is about 4 nm. **b** Autocorrelation trace and fitting curve of sech$^2$ function. Pulse duration is about 290 fs. **c** SOPs measured for filtered wavelengths of main peak and Kelly sideband. The SOP of main peak is displayed as a line. SOP of Kelly sideband is located at another point far away from that of main peak. **d** Phase diagram based on the spherical orientation angle $\theta$ and ellipticity angle $\psi$ calculated from the SOPs in **c.** Pattern is similar to that in **c.** And the variation of angle is less than 0.2 radian.

We immediately investigate the SOPs of CS buildup process to make comparison with DS. The fiber laser cavity of CS is almost the same with that of DS, but operates in an abnormal dispersion regime instead. As shown in Fig.6**a**, the averaged optical spectrum contains a main peak with central wavelength near 1566 nm and a Kelly sideband near 1561.5 nm, under the pump power of 104 mW. The 3dB bandwidth is about 4 nm. Actually, as the pump power increases to some extent, CS splits and even form soliton molecule or bound state because of wave-breaking effect at high energy, which is not discussed in this paper. In Fig.6**b**, the autocorrelation trace is measured by the autocorrelator and fitted by square of hyperbolic secant function. The full width at half maximum of CS pulses is about 450 fs. In the same way, as depicted in Fig.6**c** and **d**, we filtered the CS via the tunable optical filter and SOPs for main peak and Kelly sideband are measured by using the same polarization state analyzer. For main peak, SOPs are slowly moving along with wavelengths. The spherical orientation angle $\theta$ changes about 0.2 radian and variation of the ellipticity angle $\psi$ is nearly 0.1 radian. However, SOP of Kelly sideband differs sharply from that of main peak, which is located at another point far away.

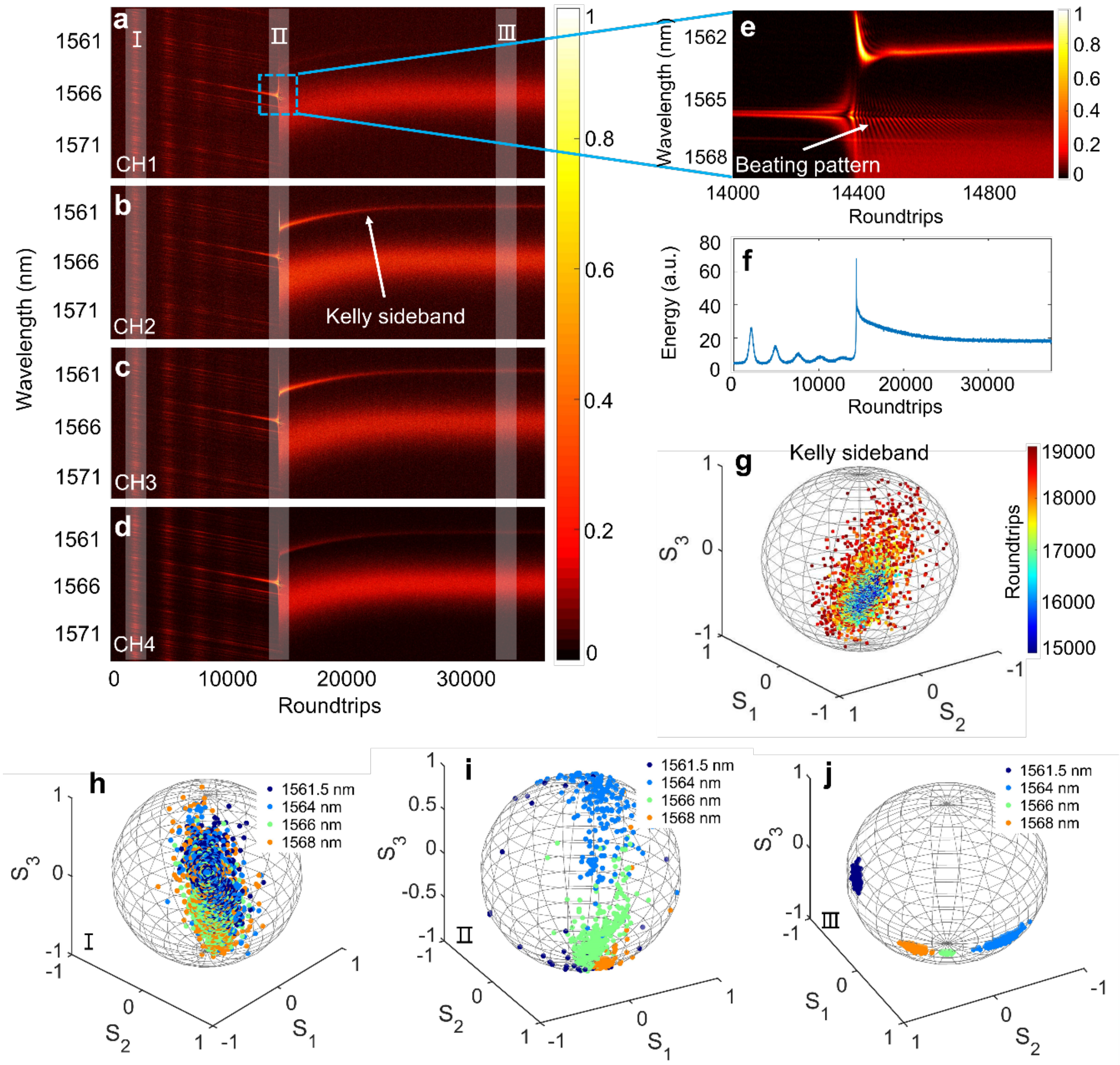

**Fig. 7. a-d** Single-shot spectra of CS build-up process of four channels under the pump power of 104 mW, up to 34000 roundtrips. Three regions I, II and III in buildup process correspond to relaxation oscillation, beating dynamics and stable mode-locking pulses are marked by white bars. **e** The beating pattern of multi pulses in region II (blue box), occurs at about $14400^{th}$ roundtrip and accompany with the present of Kelly sideband. **f** Integral energy calculated from **a**. Energy oscillation and giant peak correspond to the region I and II. In the process to region III, energy redistribution happens. **g** Temporal evolution trajectory of Kelly

sideband from 15000th roundtrip to 19000th roundtrip. With the energy redistribution, intensity of Kelly sideband varies and the SOP changes intensely. **h-j** SOPs for three regions of CSs, corresponding to white bars in **a-d**, colored by different wavelengths.

Under the circumstances, we also measure the SOPs by adopting our system. As shown in Fig.7**a-d**, the single-shot spectra of CS build-up process of four channels are detected through DFT under the pump power of 104 mW. There are more than 30000 roundtrips with time period of 34 ns are displayed. The region I, II and III are selected in buildup dynamics. For region I, relaxation oscillation starts, followed by several spikes before the region II. And frequency shift of these pulses might be induced by inelastic collision through the propagation. Integral energy calculated from **a** is shown as Fig.7**f**. In region I, energy oscillation occurs, and giant peak corresponds to the region II is mainly from dominate pulse. In the process to region III, energy redistribution happens due to the FWM process. In Fig.7**g**, the temporal evolution trace of Kelly sideband is calculated from 15000th roundtrip to 19000th roundtrip. As we can see, following the energy redistribution, intensity of Kelly sideband varies and the SOP changes intensely.

As shown in Fig.7**h**, the corresponding distribution of SOPs is disordered but overlapped, which means SOPs of these wavelengths can be in closed and even repetitive states. In region II, owing to the interference among dominate pulse and other tiny pulses caused by nonlinear refractive index and dispersion, the beating behavior occurs (Fig.7**e**). During this period, when the phase matching condition is satisfied, four-wave mixing generates new frequencies. Therefore, the spectrum is broadened followed by CS appearances and other pulses vanish.

But there is a rare blue-shift behavior with the variation of group velocity. The reason might be the energy flow from low frequency to high frequency during nonlinear process. And the corresponding SOPs begin separating in Fig.7**i**. When reaching region III, stable CSs propagate in laser cavity. As depicted in Fig.7**j**, for the stable CSs, distribution of SOPs on Poincaré sphere is similar to that of filtering result. SOPs of main peak slide along one direction, whereas that of Kelly sideband are situated in the distance, which is not the same with conventional viewpoint or hypothesis. The situation might derive from formation mechanism. CSs that circulating in fiber laser cavity experience periodic disturbance and propagate together with dispersive wave, but relative phase is varying because that only CSs are affected by nonlinearity. Despite relative phase is integer multiple of $2\pi$ at some frequencies, SOPs of Kelly sideband can be different from that of other frequencies.

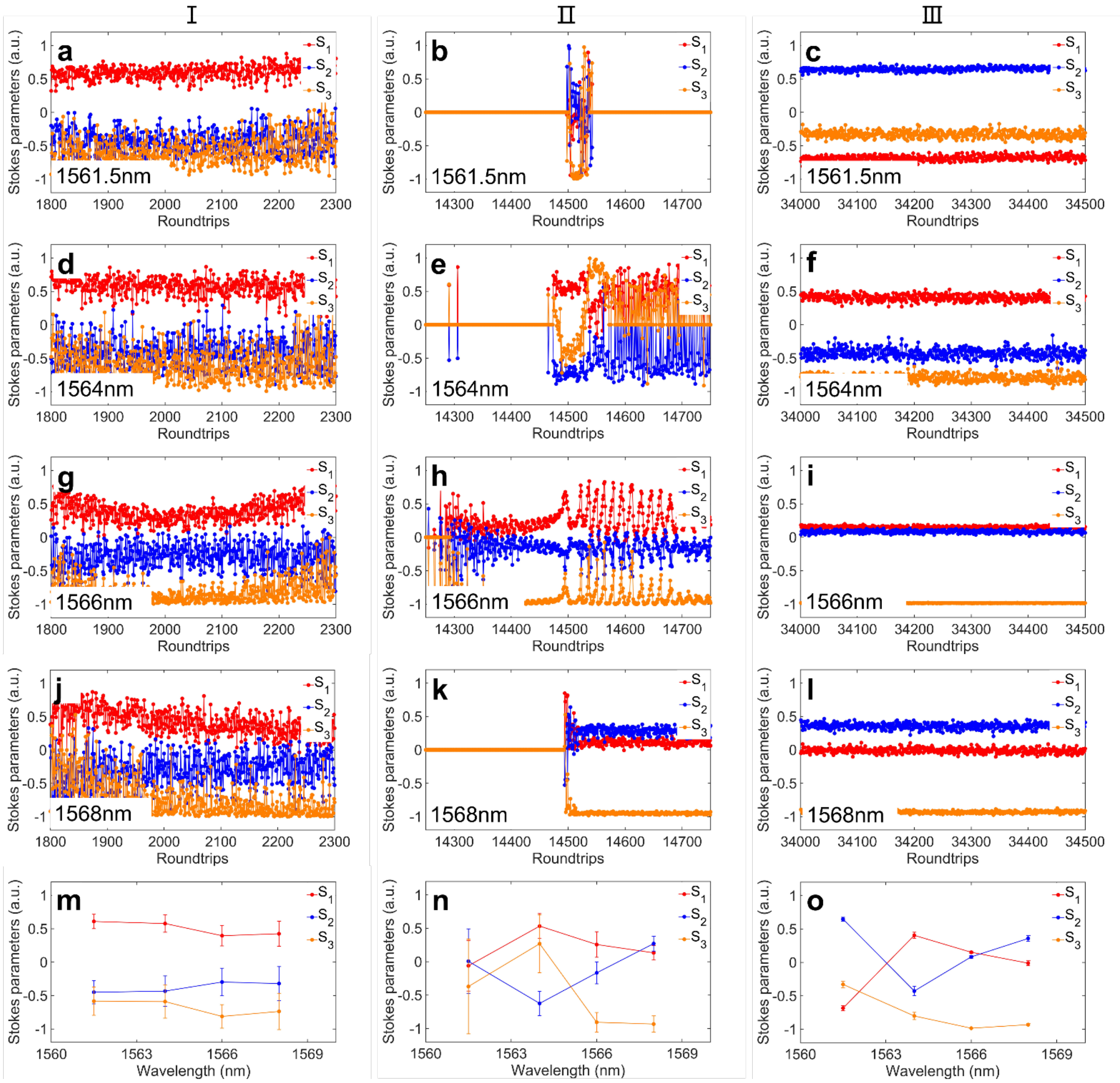

**Fig. 8. a-l** Stokes parameters of CS evolve with roundtrips for different wavelengths of 1561.5 nm, 1564 nm, 1566 nm and 1568 nm in three regions I, II and III in Fig. 7. For region I, in the process of relaxation oscillation, Stokes parameters fluctuate and glide. And the deviation is a bit large (Fig.8**m**). In region II, for the components of 1561.5 nm and 1568 nm, abrupt change of SOPs exist due to broadened spectrum in four-wave mixing. And for 1564 nm and 1566 nm, quasi-periodic-oscillation (Fig8.**e** and **h**) occurs in beating behavior in Fig.7**e**. Deviation in this region is intense because of intense oscillation as in Fig.8**n**. Stokes parameters are stabilized in region III and deviation is quite small. **m-o** Mean value and standard deviation of Stokes parameters with wavelengths in three regions.

At the same time, Stokes parameters of CS varying with roundtrips are shown in Fig.8. For region I, due to relaxation oscillation, Stokes parameters experience local fluctuations and even gliding. And the deviation has the same tendency (Fig.8**m**). In region II, for the components at the edge of spectrum, namely 1561.5 nm and 1568 nm, abrupt change of SOPs exist because of broadened spectrum in four-wave mixing. There are re-zero values in 1561.5 nm afterward, because sideband firstly locates at another place following four-wave mixing, as in region II in Fig.7. For 1564 nm and 1566 nm, quasi-periodic-oscillation corresponds to beating behavior in Fig.7**e** and the period depends on free spectral range. Deviation in this region is not flat as in region I by virtue of intense oscillation (Fig.8**n**). And in region III of mode-locking regime, Stokes parameters are stabilized at a constant level and deviation is quite small (Fig.8**o**). And Stokes parameters have a relative displacement, which coincides with gliding on Poincaré sphere in Fig.7**j**.

## Discussion

There are several parameters might limit the spectral resolution, including sampling rate of the digital oscilloscope or other electronic digitizers, analog bandwidth of photodetector and the total amount of dispersion of dispersive element. Actually, value of analog bandwidth is smaller than that of sampling rate. So the spectrum bandwidth is mainly limited by the analog bandwidth and amount of dispersion. For a higher spectrum resolution, dispersive medium with larger dispersion coefficient or longer length can be introduced in experiment. Of course, the other way is to adopt a detector with broader bandwidth. Another problem needs our attention in the frequency-to-time mapping. Each pulse is time-stretched in time domain when the pulse trains propagate through the dispersive medium and the time interval or period between consecutive pulses is determined by the laser cavity length. On this account, repetition rate of cavity itself and the amount of dispersion exerted externally should be controlled to avoid overlap between pulse trains.

Deviation among pulses attributes to two aspects. One is the laser source, single-shot pulse of which slightly differs from each other even in a stable regime. The other is the jitter or fluctuation of intensities induced by the electronic devices, particularly in measurement of high-frequency signal, precision of which is reduced in this process. This is apparent when comparing with averaging effect by optical spectrum analyzer. Of course, similar deterioration of precision can be observed in commercial polarimeter when operating in a high-speed mode.

In dissipative solitons, due to the phase shift among frequencies, including nonlinear phase shift, states of polarization among frequencies are different. When the relative phase slowly change, states of polarization constitute a smooth trajectory. Once the value of relative phase is large enough, states of polarization differ intensely, and dissipative soliton might not maintain stable in a broader spectrum range. As for buildup process of conventional solitons, states of polarization differ between Kelly sideband and soliton originating from formation mechanism. Solitons are affected by periodic disturbance and propagate together with dispersive wave, but only solitons are influenced by nonlinearity. So the relative phase between solitons and Kelly sideband is varying. SOPs of Kelly sideband can be intensely different from that of other frequencies.

In conclusion, based on division-of-amplitude, we detect four channel signals simultaneously. Combining dispersive Fourier transform, single-shot state of polarization of ultrafast lasers including dissipative solitons and conventional solitons are measured. Furthermore, we experimentally reveal the evolution of state of polarization for different stages in buildup dynamics of them. And we give some possible explanations to the dynamic process. With replacement of broadband achromatic devices in measurement system, a much broader source can be used as object of measurement. We believe this method will pave the way in research of state of polarization of ultrafast lasers in the future.

# Methods

## Adjustment of division-of-amplitude

For reducing the deviation of arrival time, we temporally measure the waveform using original femtosecond pulse laser without time-stretching, which offers more accurate time interval among four channels. And in this way, we can precisely adjust the optical path difference. Actually, the deviation is hardly avoidable by manual operation in experiment, so the residual time difference is conpensated in post-processing of data with MATLAB. The polarizers are fixed before coupling into collimators, therefore, no matter what SOP is injected into system, the output SOP of beam-splitting is always mapping into one certain SOP respectively (0°, 90°, 45° and 135°). In this case, the polarization-invariant coupling characteristic is ensured after receiving of light by collimators.

Besides, electronic noise itself also enables intensities variation, hence, when calculating the SOP from intensities in data processing, threshold value could be set to guarantee calculation from valid input light, rather than invalid electronic noise.

## Data Availability

All data used in this study are available from the corresponding author upon reasonable request.

## Code Availability

All custom codes used in this study are available from the corresponding author upon reasonable request.

## Acknowledgements

Natural Science Foundation of China (61635004, 61705023), Project supported by graduate research and innovation foundation of Chongqing, China (Grant No.CYB20061), National Postdoctoral Program for Innovative Talents (BX201600200), Postdoctoral Science Foundation of China (2017M610589), National Science Fund for Distinguished Young Scholars (61825501), the European Research Council (ERC) under the European Union’s Horizon 2020 research and innovation programme (740355), and the Russian Ministry of Science and Education (14. Y26.31.0017).